\documentclass[12pt,preprint]{aastex}

\def\gsim{\,\lower3pt\hbox{$\sim$}\llap{\raise2pt\hbox{$>$}}\,}
\def\lsim{\,\lower3pt\hbox{$\sim$}\llap{\raise2pt\hbox{$<$}}\,}
\newcommand{\be}{\begin{equation}}
\newcommand{\ee}{\end{equation}}
\newcommand{\bex}{\begin{equation}\notag}
\newcommand{\eex}{\end{equation}\notag}
\newcommand{\bea}{\begin{eqnarray}}
\newcommand{\eea}{\end{eqnarray}}
\newcommand{\beax}{\begin{eqnarray*}}
\newcommand{\eeax}{\end{eqnarray*}}
\newcommand{\ba}{\begin{array}}
\newcommand{\ea}{\end{array}}

\newcommand{\vecB}{{\mathbf B}}

\newcommand{\vecJ}{{\mathbf J}}

\begin{document}

\title{On the Mechanical Energy Available to Drive Solar Flares}
\shorttitle{Energy to Drive Flares}

\author{A.~N. McClymont, G.~H. Fisher}
\affil{Institute for Astronomy, University of Hawaii,
Honolulu, HI 96822}

\begin{abstract}
Where does solar flare energy come from?
More specifically, assuming that the ultimate source of flare energy
is mechanical energy in the convection zone,
how is this translated into energy dissipated or stored in the corona?
This question appears to have been given relatively little thought,
as attention has been focussed predominantly on mechanisms for
the rapid dissipation of coronal magnetic energy
by way of MHD instabilities and plasma micro instabilities.
We consider three types of flare theory:
the steady state ``photospheric dynamo'' model
in which flare power represents coronal dissipation of currents
generated simultaneously by sub-photospheric flows;
the ``magnetic energy storage'' model where sub-photospheric flows
again induce coronal currents but which in this case are built up
over a longer period before being released suddenly;
and ``emerging flux'' models, in which new magnetic flux rising to the
photosphere already contains free energy,
and does not require subsequent stressing by photospheric motions.
We conclude that photospheric dynamos can power only very minor flares;
that coronal energy storage can in principle
meet the requirements of a major flare,
although perhaps not the very largest flares,
but that difficulties in coupling efficiently to the energy source
may limit this mechanism to moderate sized flares;
and that emerging magnetic flux tubes, generated in the solar interior,
can carry sufficient free energy to power even the largest flares ever observed.
\end{abstract}


\section{Introduction}
\label{section:intro}
It is generally accepted that solar flares occur only in regions
of highly stressed magnetic field, and it is widely supposed that the
flare energy is stored in situ in the coronal magnetic field,
or at least that the magnetic field serves as a conduit for the energy flux
supplying the flare.
Although many mechanisms for releasing the energy stored
in stressed coronal fields have been investigated,
and models have been developed for the ways in which magnetic fields
can be stressed by motions of their photospheric footpoints,
the fundamental processes responsible for moving
the footpoints have received relatively little attention.
For instance the steady state ``photospheric dynamo'' models
assume a given photospheric flow field,
neglecting the reaction of the flow to generated $\vecJ \times \vecB$ forces.
Under this assumption, an infinite energy is available!
Here we address the question of the origin of flare power, 
a question which appears to have been considered only crudely in the past:
\cite{Stenflo1969}
concluded that kinetic energy in the convection zone
beneath a sunspot was sufficient to supply a major flare,
while \cite{Spicer1982} commented that there was probably insufficient
kinetic energy in the convection zone to directly power a flare.

In this paper,
our goal is to compute upper limits to the available energy flux
in as general a way as possible,
avoiding many of the considerations which would have to be resolved
in the construction of a realistic model.
We examine the coupling of mechanical energy in
sub-photospheric flows to the coronal magnetic field.
The flows are assumed to differ at the two footpoints of a
magnetic arch or arcade of loops,
so that the differential motion of the footpoints exerts
a twisting or shearing force on the coronal magnetic field.
We evaluate the mechanical energy of turbulent motions in
the upper convection zone and of large-scale differential rotation
as possible sources of flare energy.
We also estimate the free energy carried by a newly emerging flux tube,
generated in the solar cycle dynamo region,
which is independent of near-surface flows.

Observations suggest that magnetic flux at the photospheric level
is concentrated into tight bundles or ``flux tubes.''
The amount of mechanical energy which can be intercepted
depends on the geometry of the sub-surface magnetic flux tube,
the height and time scales over which coherent motions extend,
and the time scale for communication of forces to the solar surface.
It is believed that isolated flux tubes extend at least part way down
through the convection zone and that they are in hydrostatic equilibrium
with magnetic pressure plus internal gas pressure equal to external pressure
\citep{Fisher1989}.
We will adopt this flux tube model
and assume that flux tubes are of roughly circular cross section.
The assumptions of pressure confinement and a compact cross section
impose the main limiting factor on the power
supplied to flux tubes by convective motions.

We calculate the work done on the magnetic field in the following two limits.
If currents induced by mechanical stresses are simultaneously
dissipated resistively,
the plasma slips across the magnetic field, as in an MHD generator,
and the maximum power which can be extracted from the flow
is a fraction of the kinetic energy flux intercepted by the flux tube.
This is the ``photospheric dynamo'' model.
We assume optimum impedance matching, so that the maximum
energy flux can be extracted as flare power.

On the other hand, if resistive dissipation is negligible,
the flow distorts the magnetic field and
the work done by the gas is stored inductively.
We presume that in this ``energy storage'' model
the flow couples to the magnetic flux tube through aerodynamic drag.
The available power is computed in the same way as for the ``dynamo'' model.
To make use of this power, however, the coronal magnetic field must match the
``impedance'' (ratio of force to velocity) of the source.
We find that optimum power transfer occurs when
the opposing force due to bending of the coronal magnetic field lines
allows the flux tube to move at 1/3 of the convective driving velocity,
and that 4/27 of the kinetic energy flux can be transferred to the corona.

\begin{figure}
\includegraphics[width=5.5in]{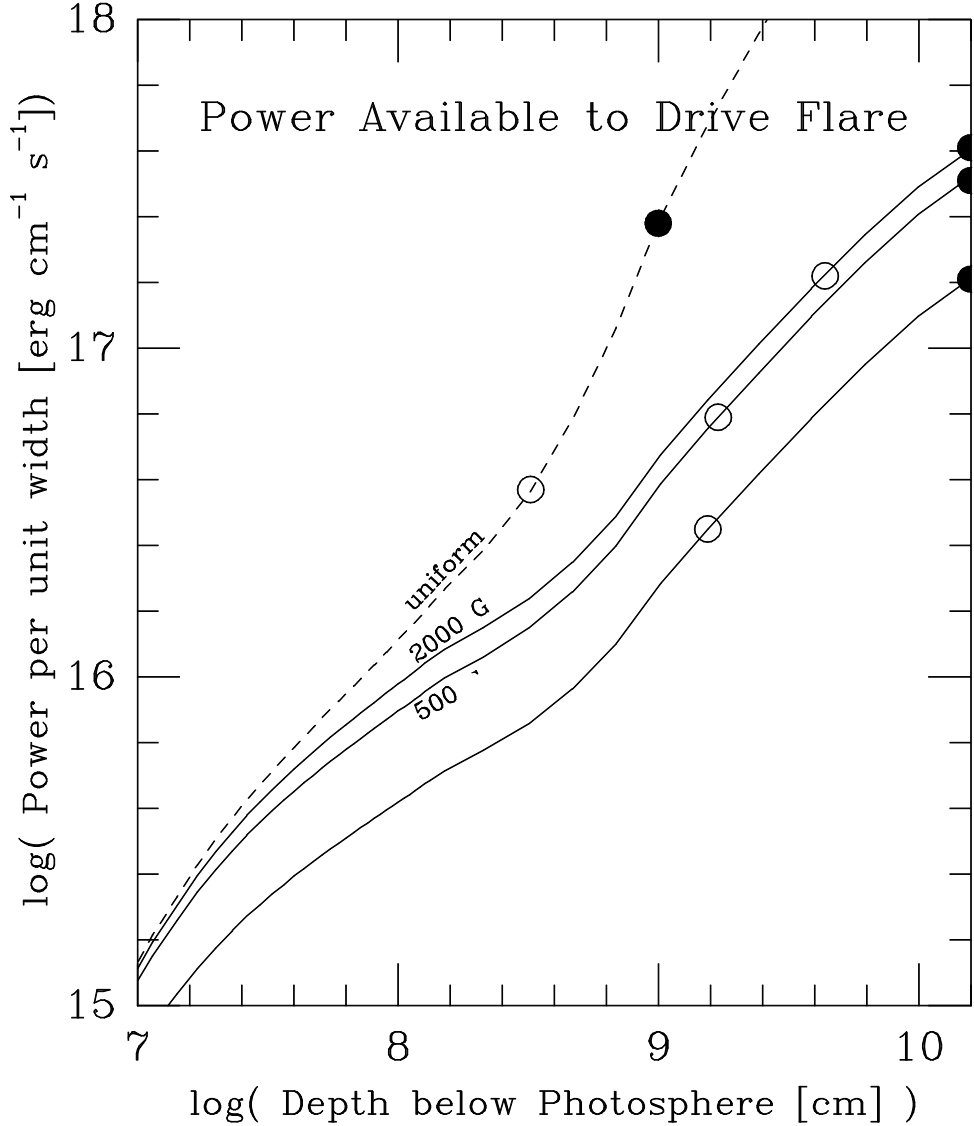}
\caption{
The kinetic energy flux of convective motions which
can be intercepted by a magnetic flux tube above depth $z$ beneath
the photosphere
for an unrealistic flux tube of uniform magnetic field strength (1000 G)
and for more realistic pressure-confined model flux tubes of
photospheric field strengths 100, 500 and 2000 G.
Open circles mark the maximum depths from which energy can propagate
on a flare time scale of 1 hour,
and filled circles mark the depths from which stresses can propagate
on a coronal energy accumulation time scale of 1 day.
(All the pressure-confined flux tubes can propagate stresses from the base of
the convection zone in 1 day.)
}
\label{figure:intro}
\end{figure}

\section{Observed Flare Energies and Time Scales}
\label{section:energies}

In a major solar flare, more than $\sim 10^{32}$ erg is dissipated
over a solar surface area of $3 \times 10^{19} {\rm \ cm}^2$
during $10^3$ - $10^4 {\rm \ s}$ (e.g., 
\citealt{Svestka1976}; \citealt{Lin1982}).
Thus the average energy flux is 
$\sim 10^9 {\rm \ erg\  cm}^{-2} {\rm \  s}^{-1}$
and the power is at least $\sim 10^{28}\ {\rm erg\ s}^{-1}$ over the flare area.
In the initial, impulsive phase of a flare, higher energy fluxes
of order $10^{10}$ - $10^{11} {\rm \ erg\ cm}^{-2} {\rm \ s}^{-1}$ are
concentrated in flare kernels of area $\sim 10^{18} {\rm \  cm}^{2}$ or less.
The impulsive phase lasts $\sim 10^{2}{\rm \  s}$
and provides the most severe test of energy release mechanisms.

Major flares are often eruptive, with much of the energy
released during mass ejections (e.g., eruptive prominences and surges).
Smaller flares (say $10^{30}$ - $10^{31}$ erg over $10^{19} {\rm \ cm}^{2}$ 
in $10^{3}{\rm \  s}$)
are frequently compact with most of the energy being carried away by radiation.
There is some evidence that the energy released impulsively during a flare
is accumulated over a period of 1 to 2 days prior to the flare.
For instance, ``homologous flares'' recur at the same location with very
similar appearance (e.g., \citealt{Svestka1976}),
and a single active region may give rise to a series of major flares
(e.g., \citealt{Lin1976}).
The requirement on power input for such energy accumulation is clearly
less severe by at least an order of magnitude
compared to the power required to drive a flare directly.

Thus the problem is to supply a power of 
$10^{27}$ - $10^{29} {\rm \ erg\  s}^{-1}$
over an area of order $10^{19} {\rm \ cm}^{2}$ 
to a directly driven ``dynamo'' flare,
or, for a ``stored energy'' flare,
to accumulate an energy surplus of order $10^{30}$ - $10^{32} {\rm \ erg}$
over a similar area,
which requires an average energy flux of 
$10^{26}$ - $10^{27} {\rm \  erg\  s}^{-1}$
over a period of order a day.

\section{Solar Flare Theories}
\label{section:theories}

In this section we briefly review the three types of flare model
under consideration in this paper.
They are the photospheric dynamo model,
in which energy of convective motions is converted directly to flare energy
on the flare time scale (minutes or hours),
the magnetic energy storage model, where pre-existing coronal fields
are stressed by convective motions over longer time scales (hours or days),
and the emerging flux model, in which new magnetic flux,
which may have accumulated free energy from convective motions
over solar cycle time scales (months or years) is able to produce flares
as soon as it reaches the solar surface.

\subsection{The Photospheric Dynamo Model}
\label{subsection:photodynamo}

This model, which is based on analogy with magnetospheric substorms,
appears in the literature in many forms (e.g., \citealt{Sen1972};
\citealt{Heyvaerts1974}; \citealt{Kan1983}; \citealt{Henoux1987}).
Its principal feature is that the flare is driven directly by the power
of photospheric flows.
All the above expositions assume that
the internal resistance of the photosphere is so high
that plasma can slip across magnetic footpoints in a steady state.
Such dissipation occurs in the weakly ionized ionosphere of the Earth
and is an important component of the theory of magnetospheric substorms.
But in the solar context, even the weakly ionized layer of the photosphere
is a good conductor.
Moreover, the weakly ionized layer of the photosphere is not ``heavy''
enough to support the required stresses,
which on a short time scale must propagate deeper into the atmosphere.
This is again contrary to the concepts developed with respect to the
Earth's magnetosphere-ionosphere system,
where the ionosphere is ``heavy,'' resistive, and in addition has an
insulating boundary beneath it.
These and other objections to the ``photospheric dynamo'' flare model
have been detailed by \cite{Melrose1987}.

Nevertheless, the basic concept of the dynamo model 
(continuous generation in the sub-photospheric region
of the power being simultaneously dissipated in a flare)
is still valid in principle.
Although the photospheric resistance is in fact very low,
such resistance must be present in the coronal part of the circuit
to account for flare dissipation.
Since the weakly ionized layer is really of no relevance,
it does not have to support the stresses,
which will be provided at a deeper level where mechanical motion
can couple more effectively to the magnetic field.

\subsection{Magnetic Energy Storage in the Corona}
\label{subsection:coronalstorage}

It is commonly believed that flares occur in regions of highly sheared
magnetic field and that the shear is produced by photospheric flows
moving the footpoints of the magnetic field (e.g., \citealt{Hagyard1986}).
Mechanisms for dissipating the field-aligned currents which result from
such twisting of the magnetic field have been proposed by,
e.g., \cite{Alfven1967} and \cite{Spicer1977a,Spicer1977b}.
We find that sub-photospheric flows can indeed move magnetic footpoints
and so stretch the coronal field.
However, simple stretching does not necessarily produce coronal currents;
production of shear requires relative rotation of the footpoints.
In our simple cylindrical model of a sub-surface flux tube,
there is clearly not enough mechanical coupling to produce rotation
unless the flux tube is fluted or fragmented beneath the photosphere,
making it easier for flows to impart torques.
Since we want to compute upper limits to the available energy flux
we leave aside such difficulties and
assume that the entire kinetic energy flux intercepted by the flux tube
can be utilized.

\subsection{Emerging Flux Models}
\label{subsection:emergingflux}

Although major flares seem to be correlated with highly sheared
magnetic fields in magnetically complex regions,
and are often associated with the eruption of pre-existing prominences,
which indicate that the magnetic configuration has existed for some time,
many flares and brightenings are observed at points of eruption of
new magnetic flux.
The energy release could be caused by either
reconnection of the new flux with overlying pre-existing magnetic field
(\citealt{Heyvaerts1977}, see also \citealt{Syrovatskii1969};
\citealt{Somov1977}; \citealt{Uchida1977}),
by conversion of the free energy due to release of the flux tube
from pressure confinement in the photosphere \citep{Spicer1986},
or by dissipation of field-aligned currents in the emerging flux tube,
as in the ``coronal energy storage'' models.
The photospheric shear flows
could be a response to the eruption of twisted flux tubes,
rather than the generator of them.

As the magnetic field of the pre-emergent flux tube is far from potential,
and the flux tube therefore carries a great amount of free energy,
it seems unnecessary to assume that the field must be stressed
by subsequent footpoint motions before a flare is produced.

\section{The Convection Zone as a Source of Flare Power}
\label{section:convectionpower}

\subsection{The Convection Zone Model}
\label{subsection:convectionmodel}

We have used the convection zone model of \cite{Spruit1974} as a basis
for estimating the available energy in convective motions.
This model, which is based on a mixing length theory,
provides an estimate of mean convection speeds as a function of depth.
These vary from $\sim 0.01$ ${\rm km\  s}^{-1}$ near the bottom of 
the convection zone,
at a depth of $10^5$ km, 
to $\sim 1$ ${\rm km \ s}^{-1}$ near the photosphere.

The energy flux available from differential rotation was computed from
the relation for the surface rotation rate,
$\Omega \ \approx \ 14.44\ -\ 3.0 \sin^2 \phi\  ^{\circ} \ {\rm day}^{-1}$,
where $\phi$ is the latitude \citep{Allen1973}.
The velocity gradient at a latitude of $\phi\ \approx\ 15 ^{\circ}$ is then
$\sim 3 \times 10^{-7}$ ${\rm s}^{-1}$.
For a magnetic loop spanning a latitude range of $\sim 3 \times 10^9$ cm,
the velocity differential between the footpoints is then of the same order
as the velocities near the bottom of the convection zone.
Since these are much smaller than the convective velocities nearer the surface,
it is clear that energy fluxes and forces similar in magnitude to those
obtainable from convective motions will only be obtained
if it is assumed that the surface rotation rate extends throughout
the convection zone,
and that flux tubes are able to convey forces from great depths.

\subsection{Models of Sub-Surface Flux Tubes}
\label{subsection:subsurfacetubes}

We have used two models of flux tubes.
In the first case,
we assume that the magnetic field strength is independent of depth,
so that the flux tube area remains constant and so intercepts
the maximum feasible fluxes of energy and momentum.
This model is not at all realistic,
and although it presents the greatest cross section at depth,
the Alfven speed, and so the ``rigidity'' of the flux tube,
drops so rapidly with increasing depth that it is unable to transmit
forces to the surface on the relevant time scales
from depths of greater than $\sim 5,000$ km.
In the second case, we use the thin flux tube approximation of 
\cite{Spruit1981}
to compute ``pressure-confined'' model flux tubes
with surface magnetic field strengths of $100 - 2000$ G,
which are rooted at the base of the convection zone
and are in hydrostatic equilibrium (see \citealt{Fisher1989}).
These flux tubes present a much smaller cross-section at depth,
but are ``stiff'' enough to transmit stresses from deep in the convection zone
to the surface in only a few hours.

\subsection{Power Available to the Photospheric Dynamo}
\label{subsection:photopower}

The maximum power available to a magnetic footpoint
down to a depth $z$ beneath the photosphere is computed from
\be
P ( z ) = \int_0^z
{ 1 \over 2} \ \rho ( z' ) \ v ( z' ) ^2 \ w ( z' ) \ v ( z' ) ~ d z'\ ,
\ee
where $\rho$ is the density, $v$ is the mean convective velocity,
and $w$ the width of the flux tube.
The results of this calculation are shown in Figure 1
for a flux tube of unit width at the photosphere.
The solid lines represent the energy flux intercepted by
pressure-confined flux tubes with surface magnetic fields of $100 - 2000$ G,
while the dashed line is for the ``uniform area'' flux tube.
On the curves open circles mark the greatest depths
from which stresses can propagate on a flare time scale of 1 hour.
(A field strength of $1000$ G was assumed for the uniform field flux tube
in this computation.)
To estimate flare power levels, we reduce the powers intercepted by 50\%,
assuming optimum impedance matching.

Thus we find that for a flux tube of surface area $3 \times 10^{19}$ 
${\rm cm}^2$ ($w\ \sim\  $ $5 \times 10^9\ {\rm cm}$), say,
the maximum power ranges from $8 \times 10^{25}$ ${\rm erg\ s}^{-1}$
for the ``uniform'' and $100$ G flux tubes
to $4 \times 10^{26}$ ${\rm erg\ s}^{-1}$ for the $2000$ G flux tube.
These energy fluxes are at least a factor of 30
too small to directly power major flares.
However,
a power level of $\sim 10^{26}$ ${\rm erg\ s}^{-1}$ 
is typical of minor subflares.
Thus the ``photospheric dynamo,'' while it might play a role
in producing the generally enhanced levels of emission in active regions,
and even small flares and brightenings,
appears to be effectively ruled out as a mechanism for major or even moderate
flares.

Such power levels, however, may be sufficient to supply pre-flare buildup
of energy over a longer period,
even without allowing for the fact that, over a longer time,
stresses may propagate from deeper in the convection zone.
This topic is discussed further in the next section.

We have also computed the power available from differential rotation,
on the assumption that the observed surface rotation rates hold all
the way to the base of the convection zone.
For a latitude span of $3 \times 10^9$ cm between footpoints,
the power available is $\lsim 10^{24}$ ${\rm erg\ s}^{-1}$ for 
all flux tube models.
Thus differential rotation certainly cannot drive flares directly,
and is very unlikely to contribute significantly to flare energy storage.

\subsection{Power Available for Coronal Energy Storage}
\label{subsection:corenstor}

Just as the ``dynamo'' model requires a matching coronal resistance to
extract maximum power from the flow,
there is an optimum ``spring constant'' for the coronal magnetic structure
in order to store magnetic energy at the highest rate.
To estimate the maximum rate at which convection zone flows can do work
on a flux tube, we simplify the problem
by assuming that a layer of the convection zone only one scale height thick,
over which the convection velocity $v$ is approximately constant,
contributes most of the power.
This assumption is consistent with our neglect of the fact that different
levels of the atmosphere may be pushing the flux tube in different directions.
The rate at which work is done on a flux tube by the drag force is then
$P \ \propto \  {1 \over 2}\  \rho ( v - V ) | v - V | V$,
where $V$ is the speed of the flux tube relative to the
other footpoint of the coronal flux tube.
Taking the other end of the coronal magnetic field line to be fixed,
$P$ is maximized when $V \ =\ 1/3 \ v$,
yielding $P_{\rm max} \approx\ (4/27) \ {1 \over 2} \rho v^3$.
Therefore in estimating the available power we have reduced the values
from Figure 1 by 4/27.

The filled circles on the curves in Figure 1 indicate the depths
from which stresses can propagate in an assumed energy
accumulation time of 24 hours.
All the pressure-confined flux tubes can transmit stresses from the base
of the convection zone in this interval.
From Figure 1, the maximum power available to any of the flux tubes of
photospheric width $5 \times 10^9$ cm (as considered in the previous section)
lies in the range $10^{26}$ ${\rm erg\ s}^{-1}$ to $3 \times 10^{26}$ 
${\rm erg\ s}^{-1}$.
When multiplied by an energy accumulation time scale of 1 day,
these yield flare energies of $10^{31}  - 3 \times 10^{31}$ erg,
typical of the energy release in a major flare (e.g., 
\citealt{Wu1986}),
although still an order of magnitude short of the energy release inferred
in the very largest flares \citep{Lin1976}.

As mentioned in the previous section,
the power levels available from differential rotation are smaller than
the above figures by about 2 orders of magnitude and so cannot explain flares.

Because the power transmitted to a flux tube must couple to the coronal
magnetic field,
it is of interest to consider the forces and displacements corresponding
to the above energy fluxes.
Convective velocities in the \cite{Spruit1974} model are $\sim 1$ 
${\rm km\ s}^{-1}$ at a
depth of 100 km beneath the photosphere,
$\sim 0.4$ ${\rm km\ s}^{-1}$ at 1000 km, 
$\sim 0.1$ ${\rm km\ s}^{-1}$ at 10,000 km,
and $\lsim 0.03$ ${\rm km\ s}^{-1}$ in the lower 
half of the convection zone,
$z \gsim$ 100,000 km.
Assuming optimum coupling with $V \ \approx\ v / 3$,
these velocities correspond respectively to footpoint displacements of
$3 \times 10^9$ cm, $10^9$ cm, $3 \times 10^8$ cm, 
and $\lsim 10^8$ cm in one day.
The latter two displacements are very small,
being much less than the width of the flux tube.

Actual observed speeds of rapidly moving magnetic features in active regions
are of order $0.2 - 0.3$ ${\rm km\ s}^{-1}$ 
(e.g., \citealt{McIntosh1981}),
which suggests that if flux tubes are indeed moved by convective motions,
they must be coupled at depths of only a few hundred kilometers below
the photosphere,
where the energy flux from convective motions falls far short of that required
for a major flare.
Our canonical flux tube would accumulate only $\sim 3 \times 10^{29}$ erg
in 24 hours, an energy typical of a small but observable subflare
(cf. 
\citealt{Wu1986}).
It is not possible, however, to dismiss rapidly moving sunspots as unconnected
with flare energy storage,
since \cite{Zirin1975},
for instance,
find that rapidly moving spots produce many flares:
flare activity ceases when the spot motion stops.

It is apparent that motions deep in the convection zone
can provide the energy of a large flare
(although possibly not the very largest)
but would move the footpoints of coronal magnetic field lines
at very slow speeds and produce very small displacements.
Thus a very ``stiff'' coronal magnetic field is required to couple
efficiently to such a driving force.
On the other hand, flows in the sub-photospheric layer,
which can energize only minor flares,
yield much smaller forces but larger velocities
and can do significant work only against much weaker coronal fields.

Consider a coronal magnetic arcade of length $L$
and width $W$ between footpoints,
with average coronal magnetic field strength $B_{CO}$.
If this structure is sheared by displacing one set of footpoints
parallel to the length $L$, by a distance $\Delta L$,
a magnetic field component $\Delta B\ =\ B_{CO} \ \Delta L / W$ is induced,
yielding an increase in the magnetic energy of
$\Delta \epsilon \ =\ B_{CO}^2 \ L \ \Delta L^2 / ( 8 \pi )$,
where we have assumed a volume $W^2 L$,
i.e., that the height of the arcade is comparable to its width.
Taking $L  \ =\   5 \times 10^9$ cm,
and using the energies and displacements computed above, we then find that
optimum coupling for a minor flare driven by shallow convection
($\Delta \epsilon\ =\ 3 \times 10^{29}$ erg, $\Delta L\  =\ 3 \times 10^9$ cm)
requires $B_{CO} \ \approx\  15$ G,
while a major flare driven by deep convection
($\Delta \epsilon\ =\ 3 \times 10^{31}$ erg, $\Delta L \ =\  3 \times 10^8$ cm)
requires $B_{CO}\  \approx\   1250$ G.

While the first case corresponds to a feasible, even low,
coronal field strength, it seems unlikely that the average coronal field
strength can be over $1000$ G.
The magnetic potential energy of such an arcade would obviously be huge,
since we are requiring a very small footpoint displacement
(and therefore very small shear)
to account for the energy of a major flare.
If such a stiff ``spring'' is not available, extraction of energy from the deep
(as opposed to upper) convection zone becomes difficult
and would require at minimum a much longer time scale for energy accumulation.
This, together with the fact that magnetic structures observed
in flaring regions appear to be highly sheared,
suggests that perhaps flare energy cannot be obtained from the lower
convection zone,
in which case we are led to the conclusion that
the ``energy storage'' model cannot account for major flares,
but only moderate ones (e.g, $3 \times 10^{30}$ erg from $z \ <\ 10^4$ km,
requiring $B_{CO} \approx\  120$ G for optimal coupling).

\subsection{Energy Available in an Erupting Flux Tube}
\label{subsection:enemergingft}

A magnetic flux tube erupting through the photosphere contains free energy
which it can give up on expanding from its pressure-confined state to a
quasi-potential configuration.
In addition to converting magnetic energy to kinetic energy of its expansion
(see 
\citealt{Spicer1986}),
it will form current sheets at the interface with pre-existing
overlying magnetic fields which can then dissipate with resulting reconnection
(see 
\citealt{Heyvaerts1977}).
Furthermore,
micro instabilities resulting in the dissipation of any field-aligned currents
it carries could be triggered by the expansion.
Thus there seems to be little reason to suppose that magnetic flux tubes
must remain quiescent until stressed by photospheric flows.

To estimate the available energy, we note that the magnetic energy in a
flux tube, $\epsilon\  \approx\ B^2 \ A \ L / 8 \pi$,
where $A$ is the cross section and $L$ the length of the flux tube,
can be written as $\epsilon\ \approx\ { \Phi^2 \ L } / { 8 \pi A }$,
where $\Phi$ is the total magnetic flux.
Now, if we assume that a flux tube,
freed from the confining pressure of the photosphere,
expands upward into the corona so that its area
increases from $A_0$ to $A_1$,
by a much larger factor than its length increases,
the change in its magnetic energy is 
$\Delta \epsilon\ \approx\ ( { \Phi^2 \ L } / { 8 \pi } )( 1 / A_1 - 1 / A_0 )$.
Assuming $A_1 \ >>\ A_0$, we have
$\Delta \epsilon \ \approx\ - B^2 \ A_0 \ L / 8 \pi$,
i.e., the bulk of the initial magnetic energy is available as free energy.
Taking values of $B_0\ = \ 1000$ G, $A_0 \ = \ 10^{19}{\rm cm}^2$
(implying a magnetic flux of $10^{22}$ Mx),
and a length of $3 \times 10^9$ cm,
we find $\sim 10^{33}$ erg of available energy.
This is more energy than we were able to obtain from convective motions
and is sufficient to supply the largest flares ever observed.

Of course the magnetic flux tube presumably originated in or near
the solar cycle dynamo region,
which probably lies at the base of the convection zone
(e.g., \citealt{Gilman1986}),
and so was subjected to roughly the same forces and convective motions
as the flux tubes considered in the previous section.
The difference is that this flux tube may have been able to accumulate energy
over a significant fraction of a solar cycle (11 years),
rather than for the 1 day considered in the previous section.

\section{Conclusions}
\label{section:conclusions}

We have considered three modes by which mechanical energy of the solar
convection zone could be converted to transient energy release in
solar flares.
The mechanical energy due to differential rotation was found
to be much too small to be significant,
as was the energy in near-photospheric convective flows.
However, the kinetic energy of motions in the upper convection zone
can power moderate flares,
while motions deep in the convection zone
appear to be able to generate the energy required by large flares.

First, we examined the ``photospheric dynamo,''
which has been proposed by a number of authors in analogy with
magnetospheric substorms and MHD generators.
Here, flare energy is dissipated in the corona simultaneously with
its generation by sub-photospheric flows across the magnetic field lines
at the footpoints of coronal magnetic fields.
We find that the kinetic energy flux in convective motions is far too small
to power flares in this way,
except perhaps for minor subflares.

Next we considered ``coronal energy storage,''
in which the kinetic energy of convective motions is intercepted
in a similar way to the photospheric dynamo.
In this case, however, the energy is stored in distortions of
the coronal magnetic field, rather than being dissipated instantaneously.
If energy is allowed to accumulate for periods of order 1 day,
this mechanism has the potential to fuel even major flares.
However, it appears to fall short by an order of magnitude
of the energy thought to be released in the very largest observed flares.
Moreover, it seems to be difficult to efficiently couple
the energy source to the coronal magnetic field
in order to extract the energy required for major flares
on time scales of order 1 day,
suggesting that convective motions may be restricted to powering
only moderate sized flares.

Last, we computed the free energy available in an ``emerging flux tube'' 
and concluded that a large flux tube should carry enough free energy
to account easily for even the largest flares.
This theory is less satisfactory than the previous two
in that the source of the free energy is
the mechanism which generates magnetic flux in the first place,
about which little is understood.
But it is the only mechanism which appears able to fully account for
the energy release in a large flare.

In conclusion,
we note that solar flares exhibit an amazing variety of phenomena,
and it is quite possible that many different forms of energization
can result in a phenomenon which we recognize as a ``flare.''
But the extreme energy requirements of a major flare narrow
the range of candidates considerably.

\acknowledgments
This work was supported by NSF
under grant ATM86-19853 and by NASA under grant NAGW86-4.


\end{document}